\documentclass[trackchanges]{aastex631}

\usepackage{amsmath}
\usepackage{booktabs}
\usepackage{footnote}
\usepackage{multirow}

\begin{document}
\title{GeV gamma-ray emission from pulsar wind nebula HESS J1356-645 with \emph{Fermi}-LAT}    

\author{Xi Liu}
\author{Xiaolei Guo}
\author{Yuliang Xin}
\author{Fengrong Zhu}
\author{Siming Liu}

\affiliation{School of Physical Science and Technology, Southwest Jiaotong University, Chengdu 610031, China; xlguo@swjtu.edu.cn}

\begin{abstract}
HESS J1356-645 is considered to be a pulsar wind nebula (PWN) associated with the pulsar PSR J1357-6429. 
We reanalyze the GeV gamma-ray emission in the direction of HESS J1356-645 with more 
than 13 years of \emph{Fermi} Large Area Telescope (LAT) data. The extended gamma-ray emission above 5 GeV is found to be spatially coincident with HESS J1356-645.
The spectrum in the energy range of 1 GeV-1 TeV can be described by a power law with an index of $\Gamma=1.51\pm0.10$. The broadband spectrum of 
HESS J1356-645 can be reproduced by a leptonic model with a broken power-law electronic spectrum. In addition, we found evidence 
that the morphology of the GeV emission from HESS J1356-645 varies with energy, a behavior which is similar to that of the PWN Vela-X. More broadband 
observations will be helpful to study the energy-dependent characteristics of HESS J1356-645.
\end{abstract}

\keywords{gamma rays: general --- ISM: individual objects (HESS J1356-645)  --- methods: data analysis}

\section{INTRODUCTION}
\par 
More than two hundred very-high-energy (VHE; $> 100$ GeV) gamma-ray sources have been detected \footnote{\url{http://tevcat2.uchicago.edu/}}, and PWN
is the largest population of the identified VHE sources. Remarkably, \cite{2021Natur.594...33C} reported the detection of ultra-high-energy
(UHE; $> 100$ TeV) gamma-ray emission from 12 Galactic gamma-ray sources. It is notable that there are one or more pulsars within most of the regions 
around the 12 UHE sources, indicating that the UHE gamma-ray emission possibly originates from PWN. PWNe are important cosmic-ray (CR) 
accelerators, and the study of gamma-ray emission from PWNe is crucial for exploring the origin of Galactic CRs.

HESS J1356-645 was first discovered by H.E.S.S. \citep{2008AIPC.1085..285R,2011A&A...533A.103H}, which is likely associated with
the pulsar PSR J1357-645. 
HESS J1356-645 is an extended gamma-ray source with $\sigma = (0.2 \pm 0.02)^\circ$ for a Gaussian model in the TeV band. The spectrum 
can be well fitted by a power law with a photon index of $\Gamma = 2.2 \pm 0.2_{\rm stat} \pm 0.2_{\rm sys}$ in the energy range of 
1 - 20 TeV. And PSR J1357-6429 is a young (characteristic age $\tau_c = 7.3$ kyr) energetic pulsar with a spin-down luminosity of 
$\dot E=3.1\times 10^{36}~\rm{erg~s^{-1}}$. The distance is estimated to be 2.5 kpc based on the dispersion measure \citep{2004ApJ...611L..25C}.
In addition, \cite{2012A&A...540A..28D} found a possible optical counterpart of PSR J1357-6429, and an extremely high transverse velocity
in the range of $1600-2000~\rm{km}~\rm{s}^{-1}$ was inferred.
However, \cite{2015MNRAS.452.3273K} gave a 90\% upper limit of $1200~\rm{km}~\rm{s}^{-1}$ with the observation of Australia 
Telescope Compact Array (ATCA) at 2.1 GHz.

In the radio band, \cite{1997MNRAS.287..722D} reported a supernova remnant (SNR) 
candidate G309.8-2.8 located only $\sim 0.1^\circ$ away from HESS J1356-645 with an extension of $15'\times 35'$ at 2.4 GHz.
The Parkes-MIT-NRAO (PMN) 4.85 GHz survey and Molonglo Galactic Plane Survey (MGPS-2) at 843 MHz also detected this SNR candidate 
with the extended structure \citep{1993AJ....105.1666G,2007yCat.8082....0M}. 
According to the flux densities, \cite{2011A&A...533A.103H} fitted the radio spectrum 
using a power law with a slope of $\alpha$ = $0.01 \pm 0.07$. Such a flat radio spectrum is much different from that of the typical shell-type SNRs,
but is within the values of indices from PWNe \citep{2009BASI...37...45G,2006ARA&A..44...17G}.
\cite{2007A&A...467L..45E} and \cite{2007ApJ...665L.143Z} carried out the X-ray observations 
with \emph{XMM-Newton} and \emph{Chandra}, and found that the spectrum of the pulsar can be described by a power-law plus blackbody model.
At the same time, \cite{2007ApJ...665L.143Z} reported a faint tail-like PWN associated with PSR J1357-6429 
using \emph{Chandra} High-Resolution Camera (HRC-S) observation, which was also confirmed by \cite{2011A&A...533A.102L} with 
\emph{Chandra} ACIS-I data. Analyses of \emph{ROSAT}/PSPC and \emph{XMM-Newton} data revealed a faint extended X-ray structure, 
which is coincident with the VHE gamma-ray emission \citep{2011A&A...533A.103H}.
Furthermore, \cite{2012ApJ...744...81C} performed deeper observations with \emph{Chandra} ACIS and \emph{XMM-Newton} EPIC data.
They reported that the X-ray emission consists of a brighter, compact PWN and a surrounding fainter, but more extended PWN.
The spectral slope of the compact component is $1.3\pm0.3$, while a slope of $1.7\pm0.2$ is obtained for the extended component.
While \cite{2015PASJ...67...43I} tried to investigate the spatial variation of the X-ray photon index from the extended
emission, no significant variation was found as a function of the distance from the pulsar.

\cite{2011A&A...533A.102L} detected the GeV gamma-ray pulsations from PSR J1357-6429, and found this emission has a spectral cutoff 
at a low energy of $\sim 800$ MeV. While \cite{2011A&A...533A.102L} and \cite{2011A&A...533A.103H} also searched for 
the GeV gamma-ray counterpart of HESS J1356-645 with {\it Fermi}-LAT, no significant GeV emission was detected. 
\cite{2013ApJ...773...77A} reported the detection of a faint GeV counterpart of HESS J1356-645 with a significance of $4.7\sigma$, 
which is insufficient to perform a more detailed analysis.

\par
In this work, we perform a complete analysis of this region with more than 13-years \emph{Fermi}-LAT data.
In Section \ref{sec:2}, we present the processes and results of data analysis.
In Section \ref{sec:3} we discuss the non-thermal radiation model, based on the multi-wavelength observations.
And the conclusion of this work is given in Section \ref{sec:4}.

\section{DATA ANALYSIS}
\label{sec:2}
\par
In this work, we used the latest Pass 8 version of \emph{Fermi}-LAT data\footnote{\url{https://fermi.gsfc.nasa.gov/ssc/data}} recorded from August 5, 2008 (Mission Elapsed Time 239587201) to January 5, 2022 (Mission Elapsed Time 663033605).
The data of ``\texttt{SOURCE}'' type (evclass=128 and evtype=3) with energy range from 5 GeV to 1 TeV was selected, 
and filtered with the recommendation of
\texttt{(DATA\underline{~}QUAL > 0) \&\& (LAT\underline{~}CONFIG == 1)}.
The region of interest (ROI) is a circle with a radius of $8^\circ$ centered at the central position of HESS J1356-645
(R.A. = $209\overset{\circ}{.}0$, decl. = $-64\overset{\circ}{.}5$).
The data were binned into 23 logarithmic energy bins, and $100\times 100$ spatial bins with a pixel of $0\overset{\circ}{.}1$
To exclude the influence of the Earth Limb, the events with zenith angle greater than $90^\circ$ were eliminated.
For the data analysis, we used the \texttt{Fermitools} version \texttt{2.0.8}\footnote{\url{https://fermi.gsfc.nasa.gov/ssc/data/analysis/software/}} to carry out the analysis,
and the instrument response function \texttt{P8R3\_SOURCE\_V3}.
The Galactic diffuse emission \texttt{gll\_iem\_v07.fits} and isotropic diffuse background \texttt{iso\_P8R3\_SOURCE\_V3\_v1.txt} provided by the Fermi Science Support Center\footnote{\url{https://fermi.gsfc.nasa.gov/ssc/data/access/lat/BackgroundModels.html}}(FSSC) were considered in the analysis. 
In addition to the two diffuse backgrounds, the model also includes the sources listed in the LAT 12-year source catalog \citep[4FGL-DR3;][]{2022arXiv220111184F} within the ROI. 
During the analysis process, the spectral parameters of sources located inside the ROI and the normalizations of the diffuse backgrounds were left free.
The standard binned likelihood method was applied to find the parameters of the sources.

\subsection{Results}
We first generated a $2^{\circ}\times 2^{\circ}$ TS map centered at HESS J1356-645, 
after subtracting the 4FGL-DR3 sources (except for HESS J1356-645) and the diffuse backgrounds, which is shown in Figure \ref{fig:1}.
Here, the TeV \citep{2011A&A...533A.103H} and 4.85 GHz radio observations of HESS J1356-645 are presented as the magenta and cyan contours, respectively.
The TS map shows that the GeV gamma-ray emission in the direction of HESS J1356-645 is in a good spatial consistence with the TeV and radio observations, suggesting a common origin.

\subsubsection{Spatial Analysis}
\label{sec:spatial}
\par
\begin{figure}[htp]
    \centering
    \includegraphics[trim={0 0.cm 0 0}, clip, width=0.48\textwidth]{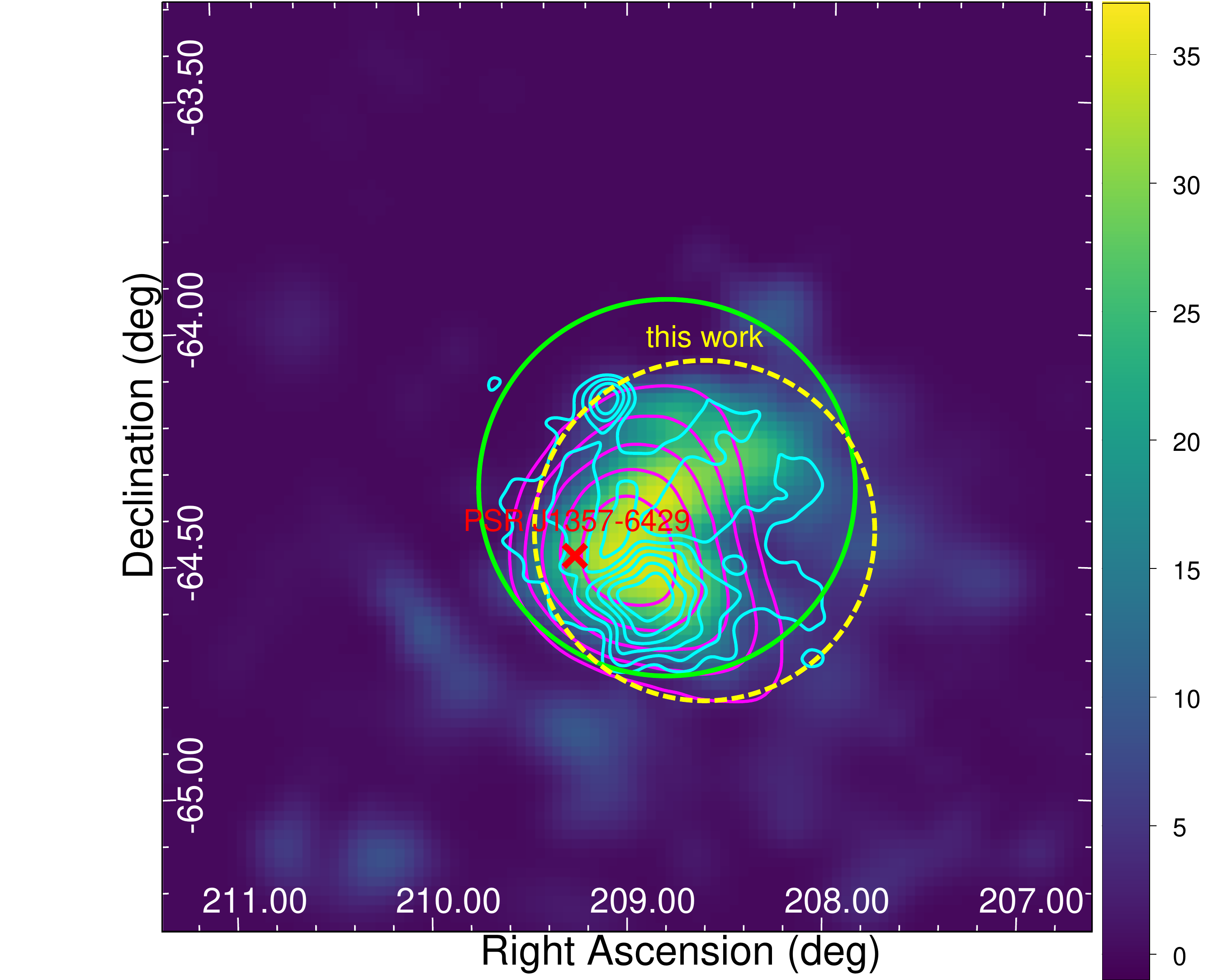}
    \caption{$2^{\circ}\times 2^{\circ}$ TS map centered on HESS J1356-645 with the data above 5 GeV, which is smoothed with a Gaussian width of $0\overset{\circ}{.}1$. 
    The yellow dashed and the green solid circles show the best-fit $r_{68}$ of the 2D Gaussian model analyzed in this work and the radius of the uniform disk template used in 4FGL-DR3, respectively.
    The magenta and cyan contours represent the H.E.S.S.} and 4.85 GHz radio observations \citep{2011A&A...533A.103H,1993AJ....105.1666G}, respectively.
    And the red cross marks the position of PSR J1357-6429 \citep{2004ApJ...611L..25C}.
    \label{fig:1}
\end{figure}

As shown in Figure \ref{fig:1}, there is a discrepancy between the GeV gamma-ray emission we detected and the 
spatial model used in 4FGL-DR3 catalog. 
Therefore, we re-analyzed the extension with an uniform disk and a two dimensional (2D) Gaussian models 
using \texttt{fermipy} \citep{2017ICRC...35..824W}. For the analysis of extension, the spectral indices of HESS J1356-645 and nearby sources
were fixed to be the best-fit values obtained in the full energy range, while the normalizations of these sources and the 
diffuse backgrounds were left free. In addition, the radio image at 4.85 GHz and the H.E.S.S. TeV image
were also tested as spatial templates.
The results of spatial analysis are given in Table \ref{tab:1}.
We found that the 2D Gaussian model with a 68\% containment radius ($r_{68}$) of 
$0\overset{\circ}{.}366$ is better to describe the GeV gamma-ray emission from HESS J1356-645, and the corresponding TS value is 109.
Based on the Akaike infotmation criterion \citep[AIC;][]{1974ITAC...19..716A}\footnote{AIC=-2ln$\mathcal{L}$+2k, where $\mathcal{L}$ is the value of maximum likelihood and k is the parameter numbers of model.}, we compared the different spatial models.
For 2D Gaussian model, the values of $\Delta\rm{AIC}$ are calculated to be 
$\Delta\rm{AIC} = \Delta\rm{AIC_{0}}-\Delta\rm{AIC_{gaus}}=7.9$ and for disk model $\Delta\rm{AIC = 4.2}$, 
where $\Delta\rm{AIC_{0}}$ represents the $\Delta$AIC value for the model of HESS J1356-645 in 4FGL.
And the values of $\Delta\rm{AIC}$ for H.E.S.S. and radio images are less than zero.
The different values of $\Delta\rm{AIC}$ suggest the Gaussian model to be the best-fit spatial template.
The $r_{68}$ of the 2D Gaussian template is plotted as a yellow dashed circle in Figure \ref{fig:1}, together with the extension in 4FGL-DR3 given by 
the analysis of Fermi-LAT extended Galactic sources \citep[FGES;][]{2017ApJ...843..139A}.
The discrepancy between them could be attributed to the different energy ranges, the improved Galactic diffuse model and more observational data used here.
In the following spectral analysis, the 2D Gaussian model is adopted as the spatial template of the GeV emission from HESS J1356-645 in order to perform a more detailed analysis.

\begin{table}[]
    \centering
    \renewcommand\arraystretch{1.5} 
    \caption{Best-fit morphological parameters of HESS J1356-645 with the data above 5 GeV.}
    \label{tab:1}
    \begin{tabular}{ccccccc}
    \hline\hline
    Spatial Model  & R.A.                                                 & decl.                                                & Extension $r_{68}$                                                           & TS &$\Delta$AIC \\ \hline
    2D Gaussian    & $208\overset{\circ}{.}614\pm 0\overset{\circ}{.}046$ & $-64\overset{\circ}{.}437\pm 0\overset{\circ}{.}046$ & $0\overset{\circ}{.}366^{+0\overset{\circ}{.}045}_{-0\overset{\circ}{.}040}$ & 109 & 7.9           \\
    Uniform Disk   & $208\overset{\circ}{.}604\pm 0\overset{\circ}{.}029$ & $-64\overset{\circ}{.}413\pm 0\overset{\circ}{.}031$ & $0\overset{\circ}{.}332^{+0\overset{\circ}{.}021}_{-0\overset{\circ}{.}021}$ & 96 & 4.2            \\ 
    H.E.S.S. Image & -                                                    & -                                                    & -                                                                            & 86 & -16.9          \\
    Radio Image    & -                                                    & -                                                    & -                                                                            & 78 & -9.3           \\ \hline
    \end{tabular}
\end{table}

\par
\begin{figure}[htp]
    \centering
    \includegraphics[trim={0 0.cm 0 0}, clip, width=0.48\textwidth]{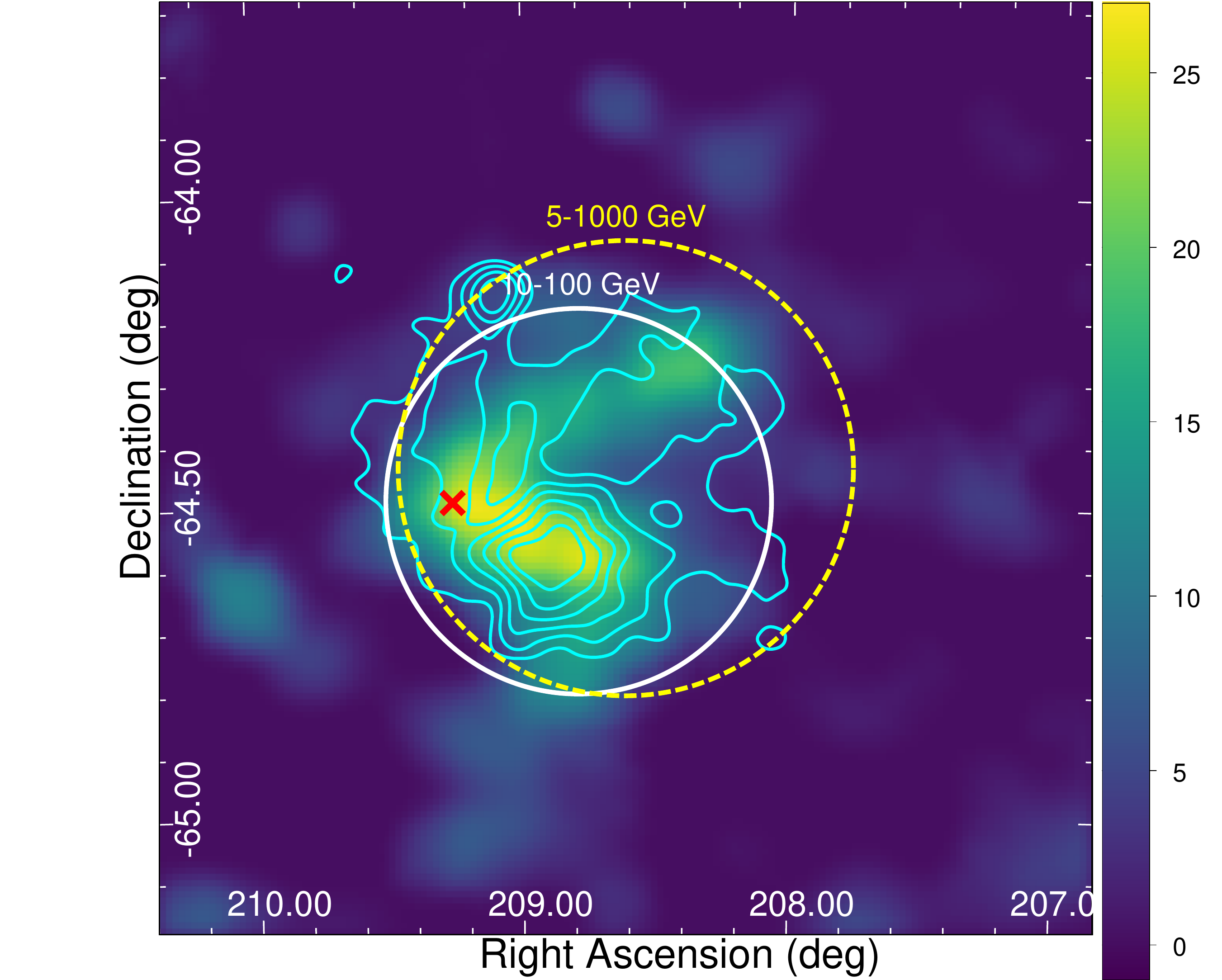}
    \includegraphics[trim={0 0.cm 0 0}, clip, width=0.48\textwidth]{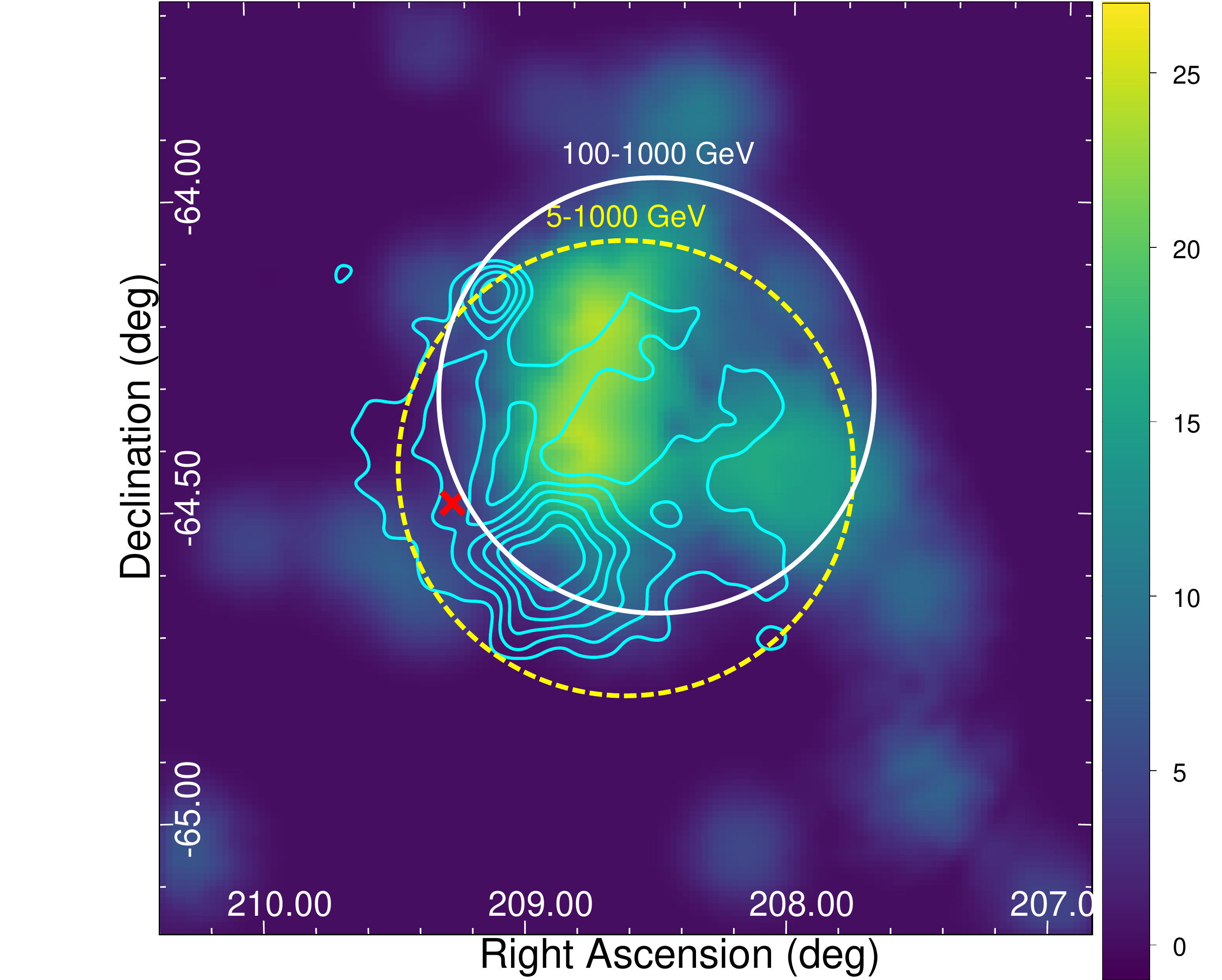}
    \caption{$1\overset{\circ}{.}5\times 1\overset{\circ}{.}5$ TS maps in the energy range of 10-100 GeV (left) and 100-1000 GeV (right). 
    The yellow dashed circle shows $r_{68}$ of HESS J1356-645 for the data above 5 GeV in this work, and the white solid circles represent
    $r_{68}$ of the Gaussian templates of the two energy bands. 
    The cyan contours represent the 4.85 GHz radio observation \citep{1993AJ....105.1666G}.
    And the red cross shows the position of PSR J1357-6429 \citep{2004ApJ...611L..25C}. 
    }
    \label{fig:2}
\end{figure}

To further explore any energy-dependent morphological characteristics of HESS J1356-645, we also performed the 
localization and extension analyses separately in the energy ranges of 10-100 GeV and 100-1000 GeV. 
And the data below 10 GeV were not included to minimise the effect of point-spread function (PSF) and diffuse background emission in the low energy band.
The analysis procedure is similar to the previous spatial analysis of the full energy band, 
and the extension is estimated by fitting a 2D Gaussian model. Table \ref{tab:2} contains the results of 
extension measurements in the two energy bands, and the corresponding TS maps are shown in Figure \ref{fig:2}. While the
extensions of both energy ranges are comparable, the central position is about $0\overset{\circ}.2$ away from each other, and both
of the central positions are offset from PSR 1357-6429. 
In addition, we noted that the GeV gamma-ray emission moves toward northwest with energy increasing, indicating the different radiation regions.

And we found that the spectrum of higher energy band is harder than that of lower energy range.
However, since the gamma-ray emission is not bright enough to perform a more detailed analysis, the statistical errors are relatively large, and the variation is possibly caused by the fluctuation of relatively weak signal or the effect of the PSF. Thus, more gamma-ray observational 
data is desired to study the energy-dependent behavior of HESS J1356-645.

\begin{table}[!htbp]
    \centering
    \renewcommand\arraystretch{1.5} 
    \caption{The results of energy-dependent extension measurements of HESS J1356-645.}
    \begin{tabular}{ccccccccc}
    \hline\hline
    Energy range     & Offset \tablenotemark{a} & Best-fit position                                                                                    & Extension $r_{68}$                                                        & TS &$\Gamma$        \\
                     &                          & (R.A., decl.)                                                                                        &                                                                           &                     \\ \hline
    10 GeV - 100 GeV & $0\overset{\circ}{.}14$  & ($208\overset{\circ}{.}79\pm0\overset{\circ}{.}05, -64\overset{\circ}{.}49\pm0\overset{\circ}{.}05$) & $0\overset{\circ}{.}31_{-0\overset{\circ}{.}05}^{+0\overset{\circ}{.}06}$ & 55 & $1.37 \pm0.27$ \\
    100 GeV - 1 TeV  & $0\overset{\circ}{.}47$  & ($208\overset{\circ}{.}50\pm0\overset{\circ}{.}07, -64\overset{\circ}{.}32\pm0\overset{\circ}{.}08$) & $0\overset{\circ}{.}35_{-0\overset{\circ}{.}06}^{+0\overset{\circ}{.}07}$ & 45 & $1.08 \pm0.41$ \\ \hline
    \end{tabular}  
    \tablenotetext{a}{The offset of the centroid of 2D-Gaussian model from the pulsar position.}
    \label{tab:2}
\end{table} 

\subsubsection{Spectral Analysis}
\par
\begin{figure}[htp]
    \centering
    \includegraphics[width=0.75\textwidth]{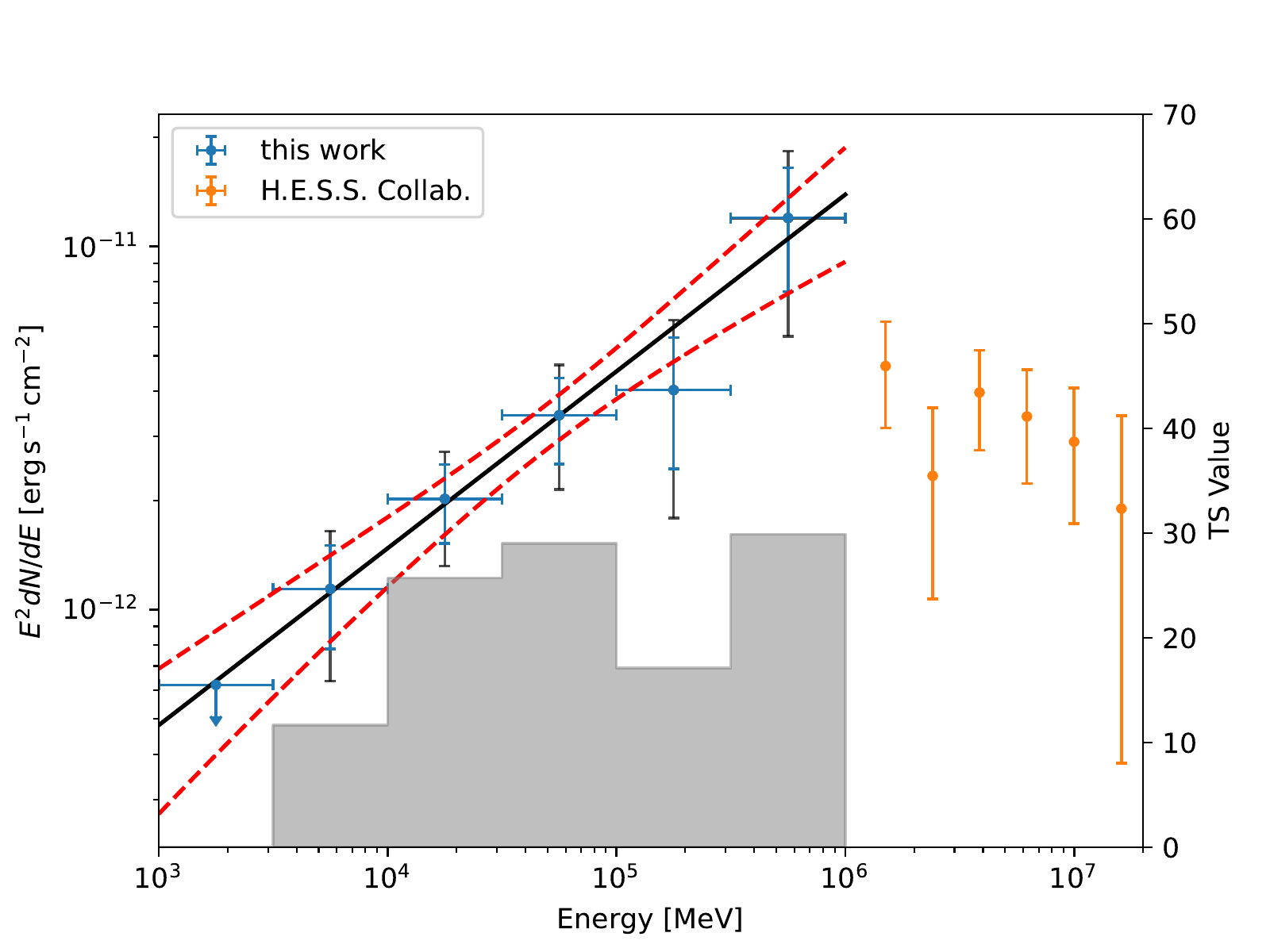}
    \caption{SED of HESS J1356-645. The blue dots are the results of {\it Fermi}-LAT data with statistical errors in this work, and the black bars mark the sums of the statistical and systematic errors considering the influence of the Galactic diffuse emission model.
    The best-fit power-law spectrum for the data above 1 GeV and the corresponding statistical error are shown as black solid and the red dashed lines, respectively.
    The TS value of each energy bin is denoted as the gray histogram.
    The VHE SED is plotted as the orange dots \citep{2011A&A...533A.103H} 
    }
    \label{fig:3}
\end{figure}

To derive the spectral energy distribution (SED), we performed the global fit from 1 GeV to 1 TeV. 
The spectrum of HESS J1356-645 can be well fitted by a power law with a spectral index of 
$\Gamma=1.51\pm0.10$ and the integrated photon flux is $(5.74\pm0.75) \times10^{-10}~\rm{photons~cm^{-2}~s^{-1}}$.
We also tested the log-parabola spectrum for HESS J1356-645, and the value of $\Delta\rm{AIC} = \rm{AIC_{Logpb}}-\rm{AIC_{PL}}$ is calculated to be 2.18, where $\rm{AIC_{Logpb}}$ and $\rm{AIC_{PL}}$ are the AIC values for log-parabola and power-law models, respectively. 
This result suggests no significant curvature for the gamma-ray spectrum of HESS J1356-645.
Then, the data were divided into 6 logarithmic energy bins. And for each energy bin, the likelihood analysis was repeated. During the analysis process,
only the normalizations of sources within the ROI and the diffuse backgrounds were set free, while the spectral indices were fixed to be the
best-fit values given in the global fit. And an upper limit of 95\% confidence level was calculated in the energy bin from 1 GeV to 3.1 GeV,
where the TS value of HESS J1356-645 is smaller than 4. 
The SED is shown in Figure \ref{fig:3}, which shows that the GeV SED could connect with the TeV spectrum smoothly.
We also estimated the systematic errors due to the Galactic diffuse emission by changing the normalization of the best-fit Galactic diffuse 
model artificially by $\pm 6\%$ as in \cite{2010ApJ...714..927A}.
The sums of statistical and systematic errors are calculated by $\sigma=\sqrt{\sigma_{\rm stat}^2+\sigma_{\rm sys}^2}$, which are shown as the black error bars in Figure \ref{fig:3}.

\section{DISCUSSION}
\label{sec:3}
\par 
Based on the results of {\em Fermi}-LAT data analysis, the extended GeV gamma-ray source is in spatial coincidence with HESS J1356-645.
And the GeV gamma-ray spectrum connects smoothly with the TeV spectrum of HESS J1356-645, 
which all support the GeV gamma-ray source as to be the low-energy counterpart of HESS J1356-645.
For the origin of the gamma-ray emission from HESS J1356-645, the SNR candidate G309.8-2.8 inside this region is considered.
However, its flat radio spectrum ($\alpha$ = $0.01 \pm 0.07$) is much different from that of shell-type SNRs, whose typical indices are
$\sim$ 0.5 \citep{2009BASI...37...45G}, but is similar to that of PWNe \citep[$-0.3 \lesssim \alpha \lesssim 0$;][]{2006ARA&A..44...17G}.
Meanwhile, the nearby pulsar PSR J1357-6429 is energetic enough to power a gamma-ray PWN, compared to other gamma-ray emitting PWNe \citep{2009ApJ...694...12M,2013ApJ...773...77A}.
Therefore, we suggest that the PWN scenario is favored for the gamma-ray emission from HESS J1356-645.

\begin{figure}
    \centering
    \includegraphics[width=0.75\textwidth]{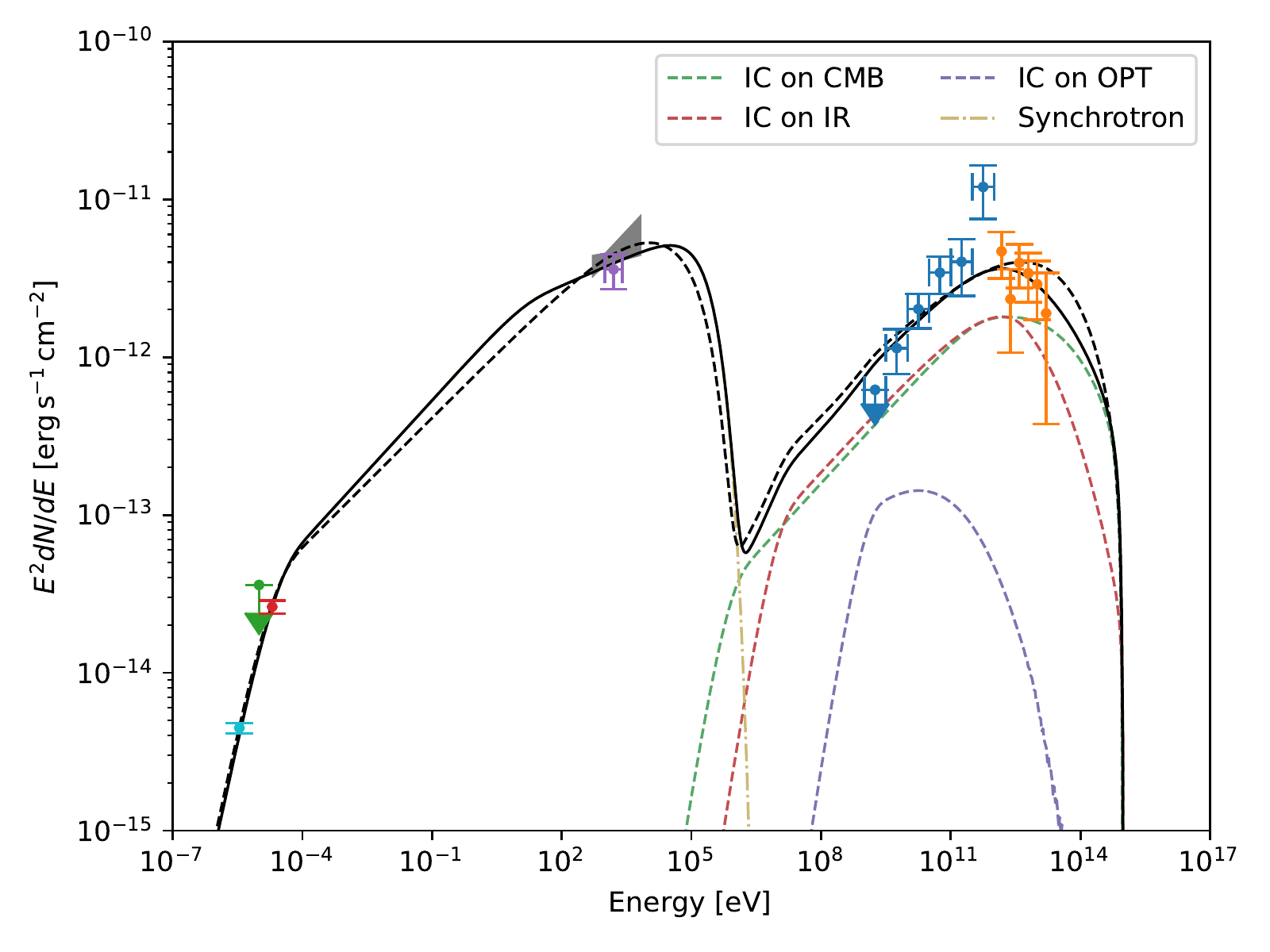}
    \caption{Leptonic model for HESS J1356-645. The broadband emission with model A and model C are represented by the black solid and dashed lines, respectively.
    The dashed lines of the different colors denote the different radiation components in model A. The radio, X-ray, and TeV gamma-ray data are taken from 
    \cite{2011A&A...533A.103H}. The radio data include the measurements from MGPS-2 at 843 MHz (cyan dot),  Parkes at 2.4 GHz (green upper limit) and PMN at 4.85 GHz (red dot). The purple dot and gray butterfly denote the X-ray observations from \emph{ROSAT}/PSPC and \emph{XMM}-Newton, respectively.}
    \label{fig:4}
\end{figure}

\begin{table}[]
    \centering
    \renewcommand\arraystretch{1.5} 
    \caption{Model parameters}
    \begin{tabular}{ccccccccc}
    \hline\hline
    & model   & $W_e$                  &  $\Gamma_1$            & $\Gamma_2$             & $E_b$                  & $E_c$               & B                      & $\chi^2$ \\
    &         & ($10^{47}$ erg)        &                        &                        & (TeV)                  & (TeV)               & ($\mu\rm{G}$)          &          \\ \hline       
    & Model A & $4.60_{-0.40}^{+0.51}$ & $2.37_{-0.06}^{+0.06}$ & $2.75_{-0.15}^{+0.19}$ & $5.39_{-0.95}^{+1.26}$ & $-$                 & $3.61_{-0.40}^{+0.59}$ & 14.6/12  \\
    & Model B & $5.08_{-0.54}^{+0.51}$ & $2.33_{-0.06}^{+0.06}$ & $2.86_{-0.18}^{+0.15}$ & $6.35_{-1.53}^{+1.61}$ & $-$                 & $3.54_{-0.45}^{+0.52}$ & 14.2/12  \\ \hline
    & Model C & $5.19_{-0.55}^{+0.66}$ & $2.45_{-0.02}^{+0.01}$ & $-$                    & $-$                    & $567_{-221}^{+384}$ & $2.92_{-0.22}^{+0.29}$ & 19.2/12  \\ \hline
    \end{tabular}
    \label{tab:3}
\end{table}

For a typical PWNe, the emission from radio to X-ray band is produced by the synchrotron radiation of relativistic electrons,
while the gamma-ray emission is generally due to the inverse-Compton scattering (ICS) of electrons (leptonic model).
To constrain the radiation mechanism of HESS J1356-645, a simple one-zone leptonic model is considered.
The electron spectrum is assumed to be a broken power law \citep[BPL;][]{2011MNRAS.410..381B},
\begin{equation}
    \notag
    \frac{dN}{dE}\propto 
    \begin{cases}
        \left(\frac{E}{E_b}\right)^{-\Gamma_1}, & E < E_{b} \\
        \left(\frac{E}{E_b}\right)^{-\Gamma_2}, & E > E_{b} \\
    \end{cases}
\end{equation}
where $E_b$ is the break energy, and $\Gamma_1$ and $\Gamma_2$ are the spectral indices in the low energy and high energy, respectively.
And the maximum energy of electrons is set to be 1 PeV, which results in the high energy cutoff.

Considering the association of PSR J1357-6429, a distance of 2.5 kpc is adopted \citep{2004ApJ...611L..25C}.
We consider two sets of background photon fields during the multi-wavelength fitting.
Model A is comprised of the cosmic microwave background (CMB), the Galactic infrared (IR; the temperature of T = 35 K and 
the energy density of $u$ = 0.66 eV $\rm{cm}^{-3}$) component and optical (T = 4600 K and $u$ = 0.94 eV $\rm{cm}^{-3}$) emission \citep{2011A&A...533A.103H}.
And in model B, the IR and optical photon fields are adopted with the values of T = 25 K \& $u$ = 0.4 eV $\rm{cm}^{-3}$ and T = 5000 K \& $u$ = 0.5 eV $\rm{cm}^{-3}$, respectively.
The multi-wavelength data are fitted by \texttt{naima} package \citep{2015ICRC...34..922Z} with the Markov Chain Monte Carlo (MCMC) algorithm. And the derived parameters are shown in Table \ref{tab:3}.

For the leptonic model with BPL spectrum, both model A and B can reproduce the multi-wavelength data, and the corresponding broadband
SED of model A is shown in Figure \ref{fig:4}. The spectral indices of electrons are 2.37 and 2.75, and the break energy is  fitted to be $\sim 5.4$ TeV. 
When we fitted the broadband spectrum, we noted that a low energy cutoff of 
$\sim 10$ GeV is needed to explain the radio data. 
Considering the flat radio spectrum ($\alpha$ = $0.01 \pm 0.07$), the spectral index below the break will be $\Gamma =2\alpha+1 \sim 1.02 $,
which is not reasonable for the particle acceleration in PWNe \citep{2015SSRv..191..519S}.
Therefore, a low energy cutoff of $\sim 10$ GeV for the electronic spectrum is considered, which is  similar to the result of \cite{2011A&A...533A.103H}.
The magnetic field strength is fitted to be $B\sim 3.7 \mu\rm{G}$, which is typical for PWNe \citep{2013ApJ...773...77A,2014JHEAp...1...31T}. 
And the total energy of electrons above 10 GeV is estimated to be $W_e\sim 4.6\times 10^{47}$ erg.
The corresponding synchrotron cooling timescale of the energy of $E_b$ is 
$\tau_{\rm{syn}} \simeq 1.56 \times 10^2 \left( \frac{E_b}{5~\rm{TeV}} \right) \left(\frac{B}{4~ \rm{\mu G}}\right)^{-2} \rm{kyr}$,
which is much larger than the characteristic age of PSR J1357-6429 ($\tau_c \sim 7.3~ \rm{kyr}$). 
In addition, the variation of spectral index of electrons predicted by synchrotron cooling is $\sim 1.0$, while the fitting gives the value of $\Delta \Gamma$ to be about 0.5. Therefore, the break energy may be intrinsic for the electron spectrum injected in PWN instead of being produced by the radiation loss \citep{2008ApJ...678L.113D}.

The electron spectrum following a power law with an exponential cutoff (PLEC; model C), 
$dN/dE \propto \left(E/E_0\right)^{-\Gamma_1} \exp\left(-E/E_c\right)$, is also tested.
The background photon fields are the same as model A, and the derived model parameters are listed in Table \ref{tab:3}. 
We found that the cutoff energy of electrons $E_c$ can not be well constrained due to the hard X-ray spectrum.
Based on the reduced $\chi^2$ values, the BPL spectrum is better to describe the broadband radiation from HESS J1356-645, which differs from the results by \cite{2011A&A...533A.103H},
and this can be attributed to the revised GeV observation data in this work.

Most of the pulsar rotational energy budget can be roughly estimated with $E_{\rm rot} = \dot E_{0} \tau_{0} (n-1)/2$ \citep{2018A&A...612A...2H}.
With the breaking index of n = 3.0 and the initial spin-down time-scale with a typical value of $\tau_{0}$ = 1 kyr, $E_{\rm rot}$ is calculated to be about 5.2 $\times 10^{48}$ erg.
Compared with the value of $\rm W_e$ in model A, the efficiency of spin-down energy converted into relativistic electrons need to be $\sim 10\%$.
It should be noted that the uncertainty of breaking index and the unknown initial spin period $P_0$ or $\tau_{0}$ of PSR J1357-6429 would result in the uncertainties of $E_{\rm rot}$ and the lepton conversion efficiency.
Furthermore, the GeV gamma-ray luminosity of HESS J1356-645 is $L_{\rm 10-1000\ GeV} \approx 1.9\times 10^{34}~(d/{\rm 2.5~kpc})^2~\rm{erg}~\rm{s}^{-1}$,
and the corresponding gamma-ray efficiency is $\eta_{\gamma}\approx0.6\%$, which is comparable with the typical PWNe efficiencies observed at GeV energies \citep{2013ApJ...773...77A}. 
The X-ray and the TeV luminosities are calculated to be
$L_{\rm X}\approx 5.9\times 10^{33}~(d/2.5~\rm{kpc})^2~erg~s^{-1}$ \citep{2015PASJ...67...43I}, 
$L_{\rm TeV}\approx 6.0\times 10^{33}~(d/2.5~\rm{kpc})^2~erg~s^{-1}$ \citep{2011A&A...533A.103H}, respectively. 
And \citet{2018A&A...612A...2H} presents the new result of $L_{\rm1-10~TeV}= (14.7\pm1.4)\times 10^{33}~(d/2.5~\rm{kpc})^2~erg~s^{-1}$ in the energy range of 1 - 10 TeV.
The ratio of $L_{\rm GeV}/L_{\rm TeV}$ is $\sim$ 3.2 and $L_{\rm GeV}/L_{\rm1-10~TeV}$ is $\sim$ 1.3, which are consistent with the mean value of $\sim 2.7^{+2.7}_{-1.4}$ for PWNe,
and the ratio of $L_{\rm GeV}/L_{\rm X} \sim 3$ is also within the range of other PWNe \citep{2013ApJ...773...77A}. 
All these characteristics strengthen the PWN scenario for HESS J1356-645.

\cite{2018ApJ...867...55X} performed a comparison of multi-wavelength data of several PWNe, and the gamma-ray spectrum of HESS J1356-645 is similar to HESS J1640-465 \citep{2014MNRAS.441.3640H,2018ApJ...867...55X}, MSH 15-52 \citep{2005A&A...435L..17A,2010ApJ...714..927A}, HESS J1825-137 \citep{2006A&A...460..365A,2011ApJ...738...42G}, and HESS J1303-631 \citep{2012A&A...548A..46H,2013ApJ...773...77A}, 
all of which have the hard GeV spectra with the gamma-ray peak at $\sim 100$ GeV. 

As mentioned in Section \ref{sec:spatial}, there is evidence that the morphology of the GeV emission from HESS J1356-645 varies with energies. 
The gamma-ray emission in lower energy band is spatially consistent with the radio nebula, while the gamma-ray emission above 100 GeV shows a different morphology,
which is similar to what is observed in the PWN Vela-X \citep{2013ApJ...774..110G,2018A&A...617A..78T}.
The energy-dependent morphologies could be produced by the different electron populations. 
The gamma rays with lower energy could be dominated by an older 
electron population accelerated in the early phase of evolution of pulsar, and the higher energy component may be from the electrons accelerated recently.
However, there is no significant difference of GeV spectral indices between the two energy ranges, which is a challenge to such a scenario.
In addition, the centroids of the gamma-ray emission in both energy ranges are offset from pulsar, and such offsets are also detected in several gamma-ray PWNe, e.g., HESS J1303-631 \citep{2012A&A...548A..46H}, Vela-X \citep{2013ApJ...774..110G}, HESS J1640-465 \citep{2018ApJ...867...55X}. 
The offset may be caused by the high proper motion of PSR J1357-6429 \citep{2012A&A...540A..28D}. 
Considering the GeV gamma-ray emission from HESS J1356-645 is not significant enough to perform a more detailed analysis, 
the future multi-band observations are desired to explore the energy-dependent behavior of HESS J1356-645.

\section{CONCLUSION}
\label{sec:4}
\par 
In this paper, we report the significant GeV gamma-ray emission from HESS J1356-645 with more than 13 years of \emph{Fermi}-LAT observation data.
The GeV gamma-ray emission from HESS J1356-645 can be described by a 2D Gaussian spatial template, and the spectrum can be fitted by a power law with 
an index of $\Gamma = 1.51 \pm 0.10$. 
The gamma-ray characteristics of HESS J1356-645 is similar to several PWNe, such as Vela-X, HESS J1825-137, HESS J1303-631 and MSH 15-52.
A leptonic model with a broken power-law spectrum can reproduce the multi-wavelength data of HESS J1356-645. 
In addition, we found the evidence that the GeV gamma-ray morphology of HESS J1356-645 varies with energies. 
And more broadband observations in future are necessary to study the energy-dependent behavior of HESS J1356-645.

\begin{acknowledgments}
\par 
We would like to thank the anonymous referee for very helpful comments, which help to improve the paper. 
This work is supported by the Natural Science Foundation for Young Scholars of Sichuan Province, China (No. 2022NSFSC1808),
the Department of Science and Technology of Sichuan Province (No. 2020YFSY0016),
the Fundamental Research Funds for the Central Universities (No. 2682021CX074, No. 2682021CX073, No. 2682022ZTPY013),
the National Natural Science Foundation of China under the grants 12103040 and 12147208,
the National Key R\&D program of China under the grant Nos. 2018YFA0404201, G2021166002L, and 2018YFA0404203, NSFC grants U1931204, 11947404 and 11761131007, DFG Sino-German Collaboration Project Nos. BU 777/15-1 and MU 4255/1-1, Department of Science.
\end{acknowledgments}

\bibliography{ref}

\begin{thebibliography}{}
\expandafter\ifx\csname natexlab\endcsname\relax\def\natexlab#1{#1}\fi
\providecommand{\url}[1]{\href{#1}{#1}}
\providecommand{\dodoi}[1]{doi:~\href{http://doi.org/#1}{\nolinkurl{#1}}}
\providecommand{\doeprint}[1]{\href{http://ascl.net/#1}{\nolinkurl{http://ascl.net/#1}}}
\providecommand{\doarXiv}[1]{\href{https://arxiv.org/abs/#1}{\nolinkurl{https://arxiv.org/abs/#1}}}

\bibitem[{{Abdo} {et~al.}(2010){Abdo}, {Ackermann}, {Ajello}, {Allafort},
  {Asano}, {Baldini}, {Ballet}, {Barbiellini}, {Baring}, {Bastieri}, {Bechtol},
  {Bellazzini}, {Berenji}, {Blandford}, {Bloom}, {Bonamente}, {Borgland},
  {Bregeon}, {Brez}, {Brigida}, {Bruel}, {Buson}, {Caliandro}, {Cameron},
  {Camilo}, {Caraveo}, {Carrigan}, {Casandjian}, {Cecchi}, {{\c{C}}elik},
  {Chekhtman}, {Cheung}, {Chiang}, {Ciprini}, {Claus}, {Cohen-Tanugi},
  {Conrad}, {den Hartog}, {Dermer}, {de Luca}, {de Palma}, {Dormody}, {Silva},
  {Drell}, {Dubois}, {Dumora}, {Farnier}, {Favuzzi}, {Fegan}, {Ferrara},
  {Focke}, {Frailis}, {Fukazawa}, {Funk}, {Fusco}, {Gargano}, {Gehrels},
  {Germani}, {Giglietto}, {Giordano}, {Glanzman}, {Godfrey}, {Gotthelf},
  {Grenier}, {Grondin}, {Grove}, {Guillemot}, {Guiriec}, {Hanabata}, {Harding},
  {Hays}, {Hobbs}, {Horan}, {Hughes}, {J{\'o}hannesson}, {Johnson}, {Johnson},
  {Johnson}, {Johnston}, {Kamae}, {Kanai}, {Kanbach}, {Katagiri}, {Kataoka},
  {Kawai}, {Keith}, {Kerr}, {Kn{\"o}dlseder}, {Kuss}, {Lande}, {Latronico},
  {Lemoine-Goumard}, {Llena Garde}, {Longo}, {Loparco}, {Lott}, {Lovellette},
  {Lubrano}, {Makeev}, {Manchester}, {Marelli}, {Mazziotta}, {McEnery},
  {Michelson}, {Mitthumsiri}, {Mizuno}, {Moiseev}, {Monte}, {Monzani},
  {Morselli}, {Moskalenko}, {Murgia}, {Nakamori}, {Nolan}, {Norris}, {Nuss},
  {Ohno}, {Ohsugi}, {Omodei}, {Orlando}, {Ormes}, {Paneque}, {Panetta},
  {Parent}, {Pelassa}, {Pepe}, {Pesce-Rollins}, {Piron}, {Porter}, {Rain{\`o}},
  {Rando}, {Razzano}, {Rea}, {Reimer}, {Reimer}, {Reposeur}, {Rodriguez},
  {Romani}, {Roth}, {Ryde}, {Sadrozinski}, {Sander}, {Saz Parkinson},
  {Sgr{\`o}}, {Siskind}, {Smith}, {Smith}, {Spandre}, {Spinelli}, {Starck},
  {Strickman}, {Suson}, {Takahashi}, {Takahashi}, {Tanaka}, {Thayer}, {Thayer},
  {Thompson}, {Thorsett}, {Tibaldo}, {Torres}, {Tosti}, {Tramacere},
  {Uchiyama}, {Usher}, {Vasileiou}, {Venter}, {Vilchez}, {Vitale}, {Waite},
  {Wang}, {Weltevrede}, {Winer}, {Wood}, {Yang}, {Ylinen}, {Ziegler}, {Fermi
  LAT Collaboration}, \& {Pulsar Timing Consortium}}]{2010ApJ...714..927A}
{Abdo}, A.~A., {Ackermann}, M., {Ajello}, M., {et~al.} 2010, \apj, 714, 927,
  \dodoi{10.1088/0004-637X/714/1/927}

\bibitem[{{Acero} {et~al.}(2013){Acero}, {Ackermann}, {Ajello}, {Allafort},
  {Baldini}, {Ballet}, {Barbiellini}, {Bastieri}, {Bechtol}, {Bellazzini},
  {Blandford}, {Bloom}, {Bonamente}, {Bottacini}, {Brandt}, {Bregeon},
  {Brigida}, {Bruel}, {Buehler}, {Buson}, {Caliandro}, {Cameron}, {Caraveo},
  {Cecchi}, {Charles}, {Chaves}, {Chekhtman}, {Chiang}, {Chiaro}, {Ciprini},
  {Claus}, {Cohen-Tanugi}, {Conrad}, {Cutini}, {Dalton}, {D'Ammando}, {de
  Palma}, {Dermer}, {Di Venere}, {Silva}, {Drell}, {Drlica-Wagner}, {Falletti},
  {Favuzzi}, {Fegan}, {Ferrara}, {Focke}, {Franckowiak}, {Fukazawa}, {Funk},
  {Fusco}, {Gargano}, {Gasparrini}, {Giglietto}, {Giordano}, {Giroletti},
  {Glanzman}, {Godfrey}, {Gr{\'e}goire}, {Grenier}, {Grondin}, {Grove},
  {Guiriec}, {Hadasch}, {Hanabata}, {Harding}, {Hayashida}, {Hayashi}, {Hays},
  {Hewitt}, {Hill}, {Horan}, {Hou}, {Hughes}, {Inoue}, {Jackson}, {Jogler},
  {J{\'o}hannesson}, {Johnson}, {Kamae}, {Kawano}, {Kerr}, {Kn{\"o}dlseder},
  {Kuss}, {Lande}, {Larsson}, {Latronico}, {Lemoine-Goumard}, {Longo},
  {Loparco}, {Lovellette}, {Lubrano}, {Marelli}, {Massaro}, {Mayer},
  {Mazziotta}, {McEnery}, {Mehault}, {Michelson}, {Mitthumsiri}, {Mizuno},
  {Monte}, {Monzani}, {Morselli}, {Moskalenko}, {Murgia}, {Nakamori}, {Nemmen},
  {Nuss}, {Ohsugi}, {Okumura}, {Orienti}, {Orlando}, {Ormes}, {Paneque},
  {Panetta}, {Perkins}, {Pesce-Rollins}, {Piron}, {Pivato}, {Porter},
  {Rain{\`o}}, {Rando}, {Razzano}, {Reimer}, {Reimer}, {Reposeur}, {Ritz},
  {Roth}, {Rousseau}, {Saz Parkinson}, {Schulz}, {Sgr{\`o}}, {Siskind},
  {Smith}, {Spandre}, {Spinelli}, {Suson}, {Takahashi}, {Takeuchi}, {Thayer},
  {Thayer}, {Thompson}, {Tibaldo}, {Tibolla}, {Tinivella}, {Torres}, {Tosti},
  {Troja}, {Uchiyama}, {Vandenbroucke}, {Vasileiou}, {Vianello}, {Vitale},
  {Werner}, {Winer}, {Wood}, \& {Yang}}]{2013ApJ...773...77A}
{Acero}, F., {Ackermann}, M., {Ajello}, M., {et~al.} 2013, \apj, 773, 77,
  \dodoi{10.1088/0004-637X/773/1/77}

\bibitem[{{Ackermann} {et~al.}(2017){Ackermann}, {Ajello}, {Baldini}, {Ballet},
  {Barbiellini}, {Bastieri}, {Bellazzini}, {Bissaldi}, {Bloom}, {Bonino},
  {Bottacini}, {Brandt}, {Bregeon}, {Bruel}, {Buehler}, {Cameron}, {Caragiulo},
  {Caraveo}, {Castro}, {Cavazzuti}, {Cecchi}, {Charles}, {Chekhtman}, {Cheung},
  {Chiaro}, {Ciprini}, {Cohen}, {Costantin}, {Costanza}, {Cutini}, {D'Ammando},
  {de Palma}, {Desiante}, {Digel}, {Di Lalla}, {Di Mauro}, {Di Venere},
  {Favuzzi}, {Fegan}, {Ferrara}, {Franckowiak}, {Fukazawa}, {Funk}, {Fusco},
  {Gargano}, {Gasparrini}, {Giglietto}, {Giordano}, {Giroletti}, {Green},
  {Grenier}, {Grondin}, {Guillemot}, {Guiriec}, {Harding}, {Hays}, {Hewitt},
  {Horan}, {Hou}, {J{\'o}hannesson}, {Kamae}, {Kuss}, {La Mura}, {Larsson},
  {Lemoine-Goumard}, {Li}, {Longo}, {Loparco}, {Lubrano}, {Magill}, {Maldera},
  {Malyshev}, {Manfreda}, {Mazziotta}, {Michelson}, {Mitthumsiri}, {Mizuno},
  {Monzani}, {Morselli}, {Moskalenko}, {Negro}, {Nuss}, {Ohsugi}, {Omodei},
  {Orienti}, {Orlando}, {Ormes}, {Paliya}, {Paneque}, {Perkins}, {Persic},
  {Pesce-Rollins}, {Petrosian}, {Piron}, {Porter}, {Principe}, {Rain{\`o}},
  {Rando}, {Razzano}, {Razzaque}, {Reimer}, {Reimer}, {Reposeur}, {Sgr{\`o}},
  {Simone}, {Siskind}, {Spada}, {Spandre}, {Spinelli}, {Suson}, {Tak},
  {Thayer}, {Thompson}, {Torres}, {Tosti}, {Troja}, {Vianello}, {Wood}, \&
  {Wood}}]{2017ApJ...843..139A}
{Ackermann}, M., {Ajello}, M., {Baldini}, L., {et~al.} 2017, \apj, 843, 139,
  \dodoi{10.3847/1538-4357/aa775a}

\bibitem[{{Aharonian} {et~al.}(2005){Aharonian}, {Akhperjanian}, {Aye},
  {Bazer-Bachi}, {Beilicke}, {Benbow}, {Berge}, {Berghaus}, {Bernl{\"o}hr},
  {Boisson}, {Bolz}, {Braun}, {Breitling}, {Brown}, {Bussons Gordo},
  {Chadwick}, {Chounet}, {Cornils}, {Costamante}, {Degrange},
  {Djannati-Ata{\"\i}}, {O'C. Drury}, {Dubus}, {Emmanoulopoulos}, {Espigat},
  {Feinstein}, {Fleury}, {Fontaine}, {Fuchs}, {Funk}, {Gallant}, {Giebels},
  {Gillessen}, {Glicenstein}, {Goret}, {Hadjichristidis}, {Hauser},
  {Heinzelmann}, {Henri}, {Hermann}, {Hinton}, {Hofmann}, {Holleran}, {Horns},
  {de Jager}, {Kh{\'e}lifi}, {Komin}, {Konopelko}, {Latham}, {Le Gallou},
  {Lemi{\`e}re}, {Lemoine-Goumard}, {Leroy}, {Lohse}, {Martineau-Huynh},
  {Marcowith}, {Masterson}, {McComb}, {de Naurois}, {Nolan}, {Noutsos},
  {Orford}, {Osborne}, {Ouchrif}, {Panter}, {Pelletier}, {Pita},
  {P{\"u}hlhofer}, {Punch}, {Raubenheimer}, {Raue}, {Raux}, {Rayner},
  {Redondo}, {Reimer}, {Reimer}, {Ripken}, {Rob}, {Rolland}, {Rowell},
  {Sahakian}, {Saug{\'e}}, {Schlenker}, {Schlickeiser}, {Schuster}, {Schwanke},
  {Siewert}, {Sol}, {Steenkamp}, {Stegmann}, {Tavernet}, {Terrier},
  {Th{\'e}oret}, {Tluczykont}, {Vasileiadis}, {Venter}, {Vincent}, {V{\"o}lk},
  \& {Wagner}}]{2005A&A...435L..17A}
{Aharonian}, F., {Akhperjanian}, A.~G., {Aye}, K.~M., {et~al.} 2005, \aap, 435,
  L17, \dodoi{10.1051/0004-6361:200500105}

\bibitem[{{Aharonian} {et~al.}(2006){Aharonian}, {Akhperjanian}, {Bazer-Bachi},
  {Beilicke}, {Benbow}, {Berge}, {Bernl{\"o}hr}, {Boisson}, {Bolz}, {Borrel},
  {Braun}, {Brown}, {B{\"u}hler}, {B{\"u}sching}, {Carrigan}, {Chadwick},
  {Chounet}, {Cornils}, {Costamante}, {Degrange}, {Dickinson},
  {Djannati-Ata{\"\i}}, {O'C. Drury}, {Dubus}, {Egberts}, {Emmanoulopoulos},
  {Espigat}, {Feinstein}, {Ferrero}, {Fiasson}, {Fontaine}, {Funk}, {Funk},
  {F{\"u}{\ss}ling}, {Gallant}, {Giebels}, {Glicenstein}, {Goret},
  {Hadjichristidis}, {Hauser}, {Hauser}, {Heinzelmann}, {Henri}, {Hermann},
  {Hinton}, {Hoffmann}, {Hofmann}, {Holleran}, {Horns}, {Jacholkowska}, {de
  Jager}, {Kendziorra}, {Kh{\'e}lifi}, {Komin}, {Konopelko}, {Kosack},
  {Latham}, {Le Gallou}, {Lemi{\`e}re}, {Lemoine-Goumard}, {Lohse}, {Martin},
  {Martineau-Huynh}, {Marcowith}, {Masterson}, {Maurin}, {McComb}, {Moulin},
  {de Naurois}, {Nedbal}, {Nolan}, {Noutsos}, {Orford}, {Osborne}, {Ouchrif},
  {Panter}, {Pelletier}, {Pita}, {P{\"u}hlhofer}, {Punch}, {Raubenheimer},
  {Raue}, {Rayner}, {Reimer}, {Reimer}, {Ripken}, {Rob}, {Rolland}, {Rowell},
  {Sahakian}, {Santangelo}, {Saug{\'e}}, {Schlenker}, {Schlickeiser},
  {Schr{\"o}der}, {Schwanke}, {Schwarzburg}, {Shalchi}, {Sol}, {Spangler},
  {Spanier}, {Steenkamp}, {Stegmann}, {Superina}, {Tavernet}, {Terrier},
  {Th{\'e}oret}, {Tluczykont}, {van Eldik}, {Vasileiadis}, {Venter}, {Vincent},
  {V{\"o}lk}, {Wagner}, \& {Ward}}]{2006A&A...460..365A}
{Aharonian}, F., {Akhperjanian}, A.~G., {Bazer-Bachi}, A.~R., {et~al.} 2006,
  \aap, 460, 365, \dodoi{10.1051/0004-6361:20065546}

\bibitem[{{Akaike}(1974)}]{1974ITAC...19..716A}
{Akaike}, H. 1974, IEEE Transactions on Automatic Control, 19, 716

\bibitem[{{Bucciantini} {et~al.}(2011){Bucciantini}, {Arons}, \&
  {Amato}}]{2011MNRAS.410..381B}
{Bucciantini}, N., {Arons}, J., \& {Amato}, E. 2011, \mnras, 410, 381,
  \dodoi{10.1111/j.1365-2966.2010.17449.x}

\bibitem[{{Camilo} {et~al.}(2004){Camilo}, {Manchester}, {Lyne}, {Gaensler},
  {Possenti}, {D'Amico}, {Stairs}, {Faulkner}, {Kramer}, {Lorimer},
  {McLaughlin}, \& {Hobbs}}]{2004ApJ...611L..25C}
{Camilo}, F., {Manchester}, R.~N., {Lyne}, A.~G., {et~al.} 2004, \apjl, 611,
  L25, \dodoi{10.1086/423620}

\bibitem[{{Cao} {et~al.}(2021){Cao}, {Aharonian}, {An}, {Axikegu}, {Bai},
  {Bao}, {Bastieri}, {Bi}, {Bi}, {Cai}, {Cai}, {Cao}, {Chang}, {Chang},
  {Chang}, {Chen}, {Chen}, {Chen}, {Chen}, {Chen}, {Chen}, {Chen}, {Chen},
  {Chen}, {Chen}, {Chen}, {Chen}, {Chen}, {Cheng}, {Cheng}, {Cui}, {Cui},
  {Cui}, {Dai}, {Dai}, {Dai}, {Danzengluobu}, {della Volpe}, {D'Ettorre
  Piazzoli}, {Dong}, {Fan}, {Fan}, {Fan}, {Fang}, {Fang}, {Feng}, {Feng},
  {Feng}, {Feng}, {Gao}, {Gao}, {Gao}, {Gao}, {Ge}, {Geng}, {Gong}, {Gou},
  {Gu}, {Guo}, {Guo}, {Guo}, {Guo}, {Han}, {He}, {He}, {He}, {He}, {He}, {He},
  {Heller}, {Hor}, {Hou}, {Hou}, {Hu}, {Hu}, {Hu}, {Hu}, {Huang}, {Huang},
  {Huang}, {Huang}, {Huang}, {Ji}, {Ji}, {Jia}, {Jiang}, {Jiang}, {Jin},
  {Kuleshov}, {Levochkin}, {Li}, {Li}, {Li}, {Li}, {Li}, {Li}, {Li}, {Li},
  {Li}, {Li}, {Li}, {Li}, {Li}, {Li}, {Li}, {Li}, {Li}, {Liang}, {Liang},
  {Lin}, {Liu}, {Liu}, {Liu}, {Liu}, {Liu}, {Liu}, {Liu}, {Liu}, {Liu}, {Liu},
  {Liu}, {Liu}, {Liu}, {Liu}, {Liu}, {Long}, {Lu}, {Lv}, {Ma}, {Ma}, {Ma},
  {Mao}, {Masood}, {Mitthumsiri}, {Montaruli}, {Nan}, {Pang},
  {Pattarakijwanich}, {Pei}, {Qi}, {Ruffolo}, {Rulev}, {S{\'a}iz}, {Shao},
  {Shchegolev}, {Sheng}, {Shi}, {Song}, {Stenkin}, {Stepanov}, {Sun}, {Sun},
  {Sun}, {Tam}, {Tang}, {Tian}, {Wang}, {Wang}, {Wang}, {Wang}, {Wang}, {Wang},
  {Wang}, {Wang}, {Wang}, {Wang}, {Wang}, {Wang}, {Wang}, {Wang}, {Wang},
  {Wang}, {Wang}, {Wang}, {Wang}, {Wang}, {Wang}, {Wei}, {Wei}, {Wei}, {Wen},
  {Wu}, {Wu}, {Wu}, {Wu}, {Wu}, {Xi}, {Xia}, {Xia}, {Xiang}, {Xiao}, {Xiao},
  {Xin}, {Xin}, {Xing}, {Xu}, {Xu}, {Xue}, {Yan}, {Yang}, {Yang}, {Yang},
  {Yang}, {Yang}, {Yang}, {Yang}, {Yao}, {Yao}, {Ye}, {Yin}, {Yin}, {You},
  {You}, {Yu}, {Yuan}, {Zeng}, {Zeng}, {Zeng}, {Zeng}, {Zha}, {Zhai}, {Zhang},
  {Zhang}, {Zhang}, {Zhang}, {Zhang}, {Zhang}, {Zhang}, {Zhang}, {Zhang},
  {Zhang}, {Zhang}, {Zhang}, {Zhang}, {Zhang}, {Zhang}, {Zhang}, {Zhang},
  {Zhang}, {Zhang}, {Zhao}, {Zhao}, {Zhao}, {Zhao}, {Zhao}, {Zheng}, {Zheng},
  {Zhou}, {Zhou}, {Zhou}, {Zhou}, {Zhou}, {Zhou}, {Zhu}, {Zhu}, {Zhu}, {Zhu},
  \& {Zuo}}]{2021Natur.594...33C}
{Cao}, Z., {Aharonian}, F.~A., {An}, Q., {et~al.} 2021, \nat, 594, 33,
  \dodoi{10.1038/s41586-021-03498-z}

\bibitem[{{Chang} {et~al.}(2012){Chang}, {Pavlov}, {Kargaltsev}, \&
  {Shibanov}}]{2012ApJ...744...81C}
{Chang}, C., {Pavlov}, G.~G., {Kargaltsev}, O., \& {Shibanov}, Y.~A. 2012,
  \apj, 744, 81, \dodoi{10.1088/0004-637X/744/2/81}

\bibitem[{{Danilenko} {et~al.}(2012){Danilenko}, {Kirichenko}, {Mennickent},
  {Pavlov}, {Shibanov}, {Zharikov}, \& {Zyuzin}}]{2012A&A...540A..28D}
{Danilenko}, A., {Kirichenko}, A., {Mennickent}, R.~E., {et~al.} 2012, \aap,
  540, A28, \dodoi{10.1051/0004-6361/201118591}

\bibitem[{{de Jager}(2008)}]{2008ApJ...678L.113D}
{de Jager}, O.~C. 2008, \apjl, 678, L113, \dodoi{10.1086/588283}

\bibitem[{{Duncan} {et~al.}(1997){Duncan}, {Stewart}, {Haynes}, \&
  {Jones}}]{1997MNRAS.287..722D}
{Duncan}, A.~R., {Stewart}, R.~T., {Haynes}, R.~F., \& {Jones}, K.~L. 1997,
  \mnras, 287, 722, \dodoi{10.1093/mnras/287.4.722}

\bibitem[{{Esposito} {et~al.}(2007){Esposito}, {Tiengo}, {de Luca}, \&
  {Mattana}}]{2007A&A...467L..45E}
{Esposito}, P., {Tiengo}, A., {de Luca}, A., \& {Mattana}, F. 2007, \aap, 467,
  L45, \dodoi{10.1051/0004-6361:20077480}

\bibitem[{{Fermi-LAT collaboration} {et~al.}(2022){Fermi-LAT collaboration},
  {:}, {Abdollahi}, {Acero}, {Baldini}, {Ballet}, {Bastieri}, {Bellazzini},
  {Berenji}, {Berretta}, {Bissaldi}, {Blandford}, {Bloom}, {Bonino}, {Brill},
  {Britto}, {Bruel}, {Burnett}, {Buson}, {Cameron}, {Caputo}, {Caraveo},
  {Castro}, {Chaty}, {Cheung}, {Chiaro}, {Cibrario}, {Ciprini},
  {Coronado-Blazquez}, {Crnogorcevic}, {Cutini}, {D'Ammando}, {De Gaetano},
  {Digel}, {Di Lalla}, {Dirirsa}, {Di Venere}, {Dominguez}, {Fallah Ramazani},
  {Fegan}, {Ferrara}, {Fiori}, {Fleischhack}, {Franckowiak}, {Fukazawa},
  {Funk}, {Fusco}, {Galanti}, {Gammaldi}, {Gargano}, {Garrappa}, {Gasparrini},
  {Giacchino}, {Giglietto}, {Giordano}, {Giroletti}, {Glanzman}, {Green},
  {Grenier}, {Grondin}, {Guillemot}, {Guiriec}, {Gustafsson}, {Harding},
  {Hays}, {Hewitt}, {Horan}, {Hou}, {Johannesson}, {Karwin}, {Kayanoki},
  {Kerr}, {Kuss}, {Landriu}, {Larsson}, {Latronico}, {Lemoine-Goumard}, {Li},
  {Liodakis}, {Longo}, {Loparco}, {Lott}, {Lubrano}, {Maldera}, {Malyshev},
  {Manfreda}, {Marti-Devesa}, {Mazziotta}, {Mereu}, {Meyer}, {Michelson},
  {Mirabal}, {Mitthumsiri}, {Mizuno}, {Moiseev}, {Monzani}, {Morselli},
  {Moskalenko}, {Negro}, {Nuss}, {Omodei}, {Orienti}, {Orlando}, {Paneque},
  {Pei}, {Perkins}, {Persic}, {Pesce-Rollins}, {Petrosian}, {Pillera}, {Poon},
  {Porter}, {Principe}, {Raino}, {Rando}, {Rani}, {Razzano}, {Razzaque},
  {Reimer}, {Reimer}, {Reposeur}, {Sanchez-Conde}, {Saz Parkinson}, {Scotton},
  {Serini}, {Sgro}, {Siskind}, {Smith}, {Spandre}, {Spinelli}, {Sueoka},
  {Suson}, {Tajima}, {Tak}, {Thayer}, {Thompson}, {Torres}, {Troja},
  {Valverde}, {Wood}, \& {Zaharijas}}]{2022arXiv220111184F}
{Fermi-LAT collaboration}, {:}, {Abdollahi}, S., {et~al.} 2022, arXiv e-prints,
  arXiv:2201.11184.
\newblock \doarXiv{2201.11184}

\bibitem[{{Gaensler} \& {Slane}(2006)}]{2006ARA&A..44...17G}
{Gaensler}, B.~M., \& {Slane}, P.~O. 2006, \araa, 44, 17,
  \dodoi{10.1146/annurev.astro.44.051905.092528}

\bibitem[{{Green}(2009)}]{2009BASI...37...45G}
{Green}, D.~A. 2009, Bulletin of the Astronomical Society of India, 37, 45.
\newblock \doarXiv{0905.3699}

\bibitem[{{Griffith} \& {Wright}(1993)}]{1993AJ....105.1666G}
{Griffith}, M.~R., \& {Wright}, A.~E. 1993, \aj, 105, 1666,
  \dodoi{10.1086/116545}

\bibitem[{{Grondin} {et~al.}(2013){Grondin}, {Romani}, {Lemoine-Goumard},
  {Guillemot}, {Harding}, \& {Reposeur}}]{2013ApJ...774..110G}
{Grondin}, M.~H., {Romani}, R.~W., {Lemoine-Goumard}, M., {et~al.} 2013, \apj,
  774, 110, \dodoi{10.1088/0004-637X/774/2/110}

\bibitem[{{Grondin} {et~al.}(2011){Grondin}, {Funk}, {Lemoine-Goumard}, {Van
  Etten}, {Hinton}, {Camilo}, {Cognard}, {Espinoza}, {Freire}, {Grove},
  {Guillemot}, {Johnston}, {Kramer}, {Lande}, {Michelson}, {Possenti},
  {Romani}, {Skilton}, {Theureau}, \& {Weltevrede}}]{2011ApJ...738...42G}
{Grondin}, M.~H., {Funk}, S., {Lemoine-Goumard}, M., {et~al.} 2011, \apj, 738,
  42, \dodoi{10.1088/0004-637X/738/1/42}

\bibitem[{{H.~E.~S.~S. Collaboration} {et~al.}(2011){H.~E.~S.~S.
  Collaboration}, {Abramowski}, {Acero}, {Aharonian}, {Akhperjanian}, {Anton},
  {Balzer}, {Barnacka}, {Barres de Almeida}, {Becherini}, {Becker}, {Behera},
  {Bernl{\"o}hr}, {Bochow}, {Boisson}, {Bolmont}, {Bordas}, {Brucker}, {Brun},
  {Brun}, {Bulik}, {B{\"u}sching}, {Carrigan}, {Casanova}, {Cerruti},
  {Chadwick}, {Charbonnier}, {Chaves}, {Cheesebrough}, {Chounet}, {Clapson},
  {Coignet}, {Cologna}, {Conrad}, {Dalton}, {Daniel}, {Davids}, {Degrange},
  {Deil}, {Dickinson}, {Djannati-Ata{\"\i}}, {Domainko}, {O'C. Drury},
  {Dubois}, {Dubus}, {Dutson}, {Dyks}, {Dyrda}, {Egberts}, {Eger}, {Espigat},
  {Fallon}, {Farnier}, {Fegan}, {Feinstein}, {Fernandes}, {Fiasson},
  {Fontaine}, {F{\"o}rster}, {F{\"u}ssling}, {Gallant}, {Gast}, {G{\'e}rard},
  {Gerbig}, {Giebels}, {Glicenstein}, {Gl{\"u}ck}, {Goret}, {G{\"o}ring},
  {H{\"a}ffner}, {Hague}, {Hampf}, {Hauser}, {Heinz}, {Heinzelmann}, {Henri},
  {Hermann}, {Hinton}, {Hoffmann}, {Hofmann}, {Hofverberg}, {Holler}, {Horns},
  {Jacholkowska}, {de Jager}, {Jahn}, {Jamrozy}, {Jung}, {Kastendieck},
  {Katarzynski}, {Katz}, {Kaufmann}, {Keogh}, {Khangulyan}, {Kh{\'e}lifi},
  {Klochkov}, {Kluzniak}, {Kneiske}, {Komin}, {Kosack}, {Kossakowski},
  {Laffon}, {Lamanna}, {Lennarz}, {Lohse}, {Lopatin}, {Lu}, {Marandon},
  {Marcowith}, {Masbou}, {Maurin}, {Maxted}, {Mayer}, {McComb}, {Medina},
  {M{\'e}hault}, {Moderski}, {Moulin}, {Naumann}, {Naumann-Godo}, {de Naurois},
  {Nedbal}, {Nekrassov}, {Nguyen}, {Nicholas}, {Niemiec}, {Nolan}, {Ohm}, {de
  Ona Wilhelmi}, {Opitz}, {Ostrowski}, {Oya}, {Panter}, {Paz Arribas},
  {Pedaletti}, {Pelletier}, {Petrucci}, {Pita}, {P{\"u}hlhofer}, {Punch},
  {Quirrenbach}, {Raue}, {Rayner}, {Reimer}, {Reimer}, {Renaud}, {de Los
  Reyes}, {Rieger}, {Ripken}, {Rob}, {Rosier-Lees}, {Rowell}, {Rudak},
  {Rulten}, {Ruppel}, {Sahakian}, {Sanchez}, {Santangelo}, {Schlickeiser},
  {Sch{\"o}ck}, {Schulz}, {Schwanke}, {Schwarzburg}, {Schwemmer}, {Sikora},
  {Skilton}, {Sol}, {Spengler}, {Stawarz}, {Steenkamp}, {Stegmann}, {Stinzing},
  {Stycz}, {Sushch}, {Szostek}, {Tavernet}, {Terrier}, {Tluczykont},
  {Valerius}, {van Eldik}, {Vasileiadis}, {Venter}, {Vialle}, {Viana},
  {Vincent}, {V{\"o}lk}, {Volpe}, {Vorobiov}, {Vorster}, {Wagner}, {Ward},
  {White}, {Wierzcholska}, {Zacharias}, {Zajczyk}, {Zdziarski}, {Zech}, \&
  {Zechlin}}]{2011A&A...533A.103H}
{H.~E.~S.~S. Collaboration}, {Abramowski}, A., {Acero}, F., {et~al.} 2011,
  \aap, 533, A103, \dodoi{10.1051/0004-6361/201117445}

\bibitem[{{H.~E.~S.~S. Collaboration} {et~al.}(2012){H.~E.~S.~S.
  Collaboration}, {Abramowski}, {Acero}, {Aharonian}, {Akhperjanian}, {Anton},
  {Balenderan}, {Balzer}, {Barnacka}, {Becherini}, {Becker}, {Bernl{\"o}hr},
  {Birsin}, {Biteau}, {Bochow}, {Boisson}, {Bolmont}, {Bordas}, {Brucker},
  {Brun}, {Brun}, {Bulik}, {B{\"u}sching}, {Carrigan}, {Casanova}, {Cerruti},
  {Chadwick}, {Charbonnier}, {Chaves}, {Cheesebrough}, {Cologna}, {Conrad},
  {Couturier}, {Dalton}, {Daniel}, {Davids}, {Degrange}, {Deil}, {Dickinson},
  {Djannati-Ata{\"\i}}, {Domainko}, {Drury}, {Dubus}, {Dutson}, {Dyks},
  {Dyrda}, {Egberts}, {Eger}, {Espigat}, {Fallon}, {Farnier}, {Fegan},
  {Feinstein}, {Fernandes}, {Fiasson}, {Fontaine}, {F{\"o}rster},
  {F{\"u}{\ss}ling}, {Gajdus}, {Gallant}, {Garrigoux}, {Gast}, {G{\'e}rard},
  {Giebels}, {Glicenstein}, {Gl{\"u}ck}, {G{\"o}ring}, {Grondin},
  {H{\"a}ffner}, {Hague}, {Hahn}, {Hampf}, {Harris}, {Hauser}, {Heinz},
  {Heinzelmann}, {Henri}, {Hermann}, {Hillert}, {Hinton}, {Hofmann},
  {Hofverberg}, {Holler}, {Horns}, {Jacholkowska}, {Jahn}, {Jamrozy}, {Jung},
  {Kastendieck}, {Katarzy{\'n}ski}, {Katz}, {Kaufmann}, {Kh{\'e}lifi},
  {Klochkov}, {Klu{\'z}niak}, {Kneiske}, {Komin}, {Kosack}, {Kossakowski},
  {Krayzel}, {Laffon}, {Lamanna}, {Lenain}, {Lennarz}, {Lohse}, {Lopatin},
  {Lu}, {Marandon}, {Marcowith}, {Masbou}, {Maurin}, {Maxted}, {Mayer},
  {McComb}, {Medina}, {M{\'e}hault}, {Menzler}, {Moderski}, {Mohamed},
  {Moulin}, {Naumann}, {Naumann-Godo}, {de Naurois}, {Nedbal}, {Nekrassov},
  {Nguyen}, {Nicholas}, {Niemiec}, {Nolan}, {Ohm}, {de O{\~n}a Wilhelmi},
  {Opitz}, {Ostrowski}, {Oya}, {Panter}, {Paz Arribas}, {Pekeur}, {Pelletier},
  {Perez}, {Petrucci}, {Peyaud}, {Pita}, {P{\"u}hlhofer}, {Punch},
  {Quirrenbach}, {Raue}, {Reimer}, {Reimer}, {Renaud}, {de los Reyes},
  {Rieger}, {Ripken}, {Rob}, {Rosier-Lees}, {Rowell}, {Rudak}, {Rulten},
  {Sahakian}, {Sanchez}, {Santangelo}, {Schlickeiser}, {Schulz}, {Schwanke},
  {Schwarzburg}, {Schwemmer}, {Sheidaei}, {Skilton}, {Sol}, {Spengler},
  {Stawarz}, {Steenkamp}, {Stegmann}, {Stinzing}, {Stycz}, {Sushch}, {Szostek},
  {Tavernet}, {Terrier}, {Tluczykont}, {Valerius}, {van Eldik}, {Vasileiadis},
  {Venter}, {Viana}, {Vincent}, {V{\"o}lk}, {Volpe}, {Vorobiov}, {Vorster},
  {Wagner}, {Ward}, {White}, {Wierzcholska}, {Zacharias}, {Zajczyk},
  {Zdziarski}, {Zech}, \& {Zechlin}}]{2012A&A...548A..46H}
---. 2012, \aap, 548, A46, \dodoi{10.1051/0004-6361/201219814}

\bibitem[{{H.~E.~S.~S. Collaboration} {et~al.}(2014){H.~E.~S.~S.
  Collaboration}, {Abramowski}, {Aharonian}, {Ait Benkhali}, {Akhperjanian},
  {Ang{\"u}ner}, {Anton}, {Balenderan}, {Balzer}, {Barnacka}, {Becherini},
  {Becker Tjus}, {Bernl{\"o}hr}, {Birsin}, {Bissaldi}, {Biteau},
  {B{\"o}ttcher}, {Boisson}, {Bolmont}, {Bordas}, {Brucker}, {Brun}, {Brun},
  {Bulik}, {Carrigan}, {Casanova}, {Cerruti}, {Chadwick}, {Chalme-Calvet},
  {Chaves}, {Cheesebrough}, {Chr{\'e}tien}, {Colafrancesco}, {Cologna},
  {Conrad}, {Couturier}, {Cui}, {Dalton}, {Daniel}, {Davids}, {Degrange},
  {Deil}, {deWilt}, {Dickinson}, {Djannati-Ata{\"\i}}, {Domainko}, {Drury},
  {Dubus}, {Dutson}, {Dyks}, {Dyrda}, {Edwards}, {Egberts}, {Eger}, {Espigat},
  {Farnier}, {Fegan}, {Feinstein}, {Fernandes}, {Fernandez}, {Fiasson},
  {Fontaine}, {F{\"o}rster}, {F{\"u}{\ss}ling}, {Gajdus}, {Gallant},
  {Garrigoux}, {Giavitto}, {Giebels}, {Glicenstein}, {Grondin},
  {Grudzi{\'n}ska}, {H{\"a}ffner}, {Hahn}, {Harris}, {Heinzelmann}, {Henri},
  {Hermann}, {Hervet}, {Hillert}, {Hinton}, {Hofmann}, {Hofverberg}, {Holler},
  {Horns}, {Jacholkowska}, {Jahn}, {Jamrozy}, {Janiak}, {Jankowsky}, {Jung},
  {Kastendieck}, {Katarzy{\'n}ski}, {Katz}, {Kaufmann}, {Kh{\'e}lifi},
  {Kieffer}, {Klepser}, {Klochkov}, {Klu{\'z}niak}, {Kneiske}, {Kolitzus},
  {Komin}, {Kosack}, {Krakau}, {Krayzel}, {Kr{\"u}ger}, {Laffon}, {Lamanna},
  {Lefaucheur}, {Lemi{\`e}re}, {Lemoine-Goumard}, {Lenain}, {Lennarz}, {Lohse},
  {Lopatin}, {Lu}, {Marandon}, {Marcowith}, {Marx}, {Maurin}, {Maxted},
  {Mayer}, {McComb}, {M{\'e}hault}, {Meintjes}, {Menzler}, {Meyer}, {Moderski},
  {Mohamed}, {Moulin}, {Murach}, {Naumann}, {de Naurois}, {Niemiec}, {Nolan},
  {Oakes}, {Ohm}, {de O{\~n}a Wilhelmi}, {Opitz}, {Ostrowski}, {Oya}, {Panter},
  {Parsons}, {Arribas}, {Pekeur}, {Pelletier}, {Perez}, {Petrucci}, {Peyaud},
  {Pita}, {Poon}, {P{\"u}hlhofer}, {Punch}, {Quirrenbach}, {Raab}, {Raue},
  {Reimer}, {Reimer}, {Renaud}, {Reyes}, {Rieger}, {Rob}, {Romoli},
  {Rosier-Lees}, {Rowell}, {Rudak}, {Rulten}, {Sahakian}, {Sanchez},
  {Santangelo}, {Schlickeiser}, {Sch{\"u}ssler}, {Schulz}, {Schwanke},
  {Schwarzburg}, {Schwemmer}, {Sol}, {Spengler}, {Spies}, {Stawarz},
  {Steenkamp}, {Stegmann}, {Stinzing}, {Stycz}, {Sushch}, {Szostek},
  {Tavernet}, {Tavernier}, {Taylor}, {Terrier}, {Tluczykont}, {Trichard},
  {Valerius}, {van Eldik}, {van Soelen}, {Vasileiadis}, {Venter}, {Viana},
  {Vincent}, {Vink}, {V{\"o}lk}, {Volpe}, {Vorster}, {Vuillaume}, {Wagner},
  {Wagner}, {Ward}, {Weidinger}, {Weitzel}, {White}, {Wierzcholska},
  {Willmann}, {W{\"o}rnlein}, {Wouters}, {Zabalza}, {Zacharias}, {Zajczyk},
  {Zdziarski}, {Zech}, \& {Zechlin}}]{2014MNRAS.441.3640H}
{H.~E.~S.~S. Collaboration}, {Abramowski}, A., {Aharonian}, F., {et~al.} 2014,
  \mnras, 441, 3640, \dodoi{10.1093/mnras/stu826}

\bibitem[{{H.~E.~S.~S. Collaboration} {et~al.}(2018){H.~E.~S.~S.
  Collaboration}, {Abdalla}, {Abramowski}, {Aharonian}, {Ait Benkhali},
  {Akhperjanian}, {Andersson}, {Ang{\"u}ner}, {Arrieta}, {Aubert}, {Backes},
  {Balzer}, {Barnard}, {Becherini}, {Becker Tjus}, {Berge}, {Bernhard},
  {Bernl{\"o}hr}, {Blackwell}, {B{\"o}ttcher}, {Boisson}, {Bolmont}, {Bordas},
  {Bregeon}, {Brun}, {Brun}, {Bryan}, {Bulik}, {Capasso}, {Carr}, {Carrigan},
  {Casanova}, {Cerruti}, {Chakraborty}, {Chalme-Calvet}, {Chaves}, {Chen},
  {Chevalier}, {Chr{\'e}tien}, {Colafrancesco}, {Cologna}, {Condon}, {Conrad},
  {Couturier}, {Cui}, {Davids}, {Degrange}, {Deil}, {Devin}, {deWilt},
  {Dirson}, {Djannati-Ata{\"\i}}, {Domainko}, {Donath}, {Drury}, {Dubus},
  {Dutson}, {Dyks}, {Edwards}, {Egberts}, {Eger}, {Ernenwein}, {Eschbach},
  {Farnier}, {Fegan}, {Fernandes}, {Fiasson}, {Fontaine}, {F{\"o}rster},
  {Funk}, {F{\"u}{\ss}ling}, {Gabici}, {Gajdus}, {Gallant}, {Garrigoux},
  {Giavitto}, {Giebels}, {Glicenstein}, {Gottschall}, {Goyal}, {Grondin},
  {Hadasch}, {Hahn}, {Haupt}, {Hawkes}, {Heinzelmann}, {Henri}, {Hermann},
  {Hervet}, {Hillert}, {Hinton}, {Hofmann}, {Hoischen}, {Holler}, {Horns},
  {Ivascenko}, {Jacholkowska}, {Jamrozy}, {Janiak}, {Jankowsky}, {Jankowsky},
  {Jingo}, {Jogler}, {Jouvin}, {Jung-Richardt}, {Kastendieck},
  {Katarzy{\'n}ski}, {Katz}, {Kerszberg}, {Kh{\'e}lifi}, {Kieffer}, {King},
  {Klepser}, {Klochkov}, {Klu{\'z}niak}, {Kolitzus}, {Komin}, {Kosack},
  {Krakau}, {Kraus}, {Krayzel}, {Kr{\"u}ger}, {Laffon}, {Lamanna}, {Lau},
  {Lees}, {Lefaucheur}, {Lefranc}, {Lemi{\`e}re}, {Lemoine-Goumard}, {Lenain},
  {Leser}, {Lohse}, {Lorentz}, {Liu}, {L{\'o}pez-Coto}, {Lypova}, {Marandon},
  {Marcowith}, {Mariaud}, {Marx}, {Maurin}, {Maxted}, {Mayer}, {Meintjes},
  {Meyer}, {Mitchell}, {Moderski}, {Mohamed}, {Mohrmann}, {Mor{\r{a}}},
  {Moulin}, {Murach}, {de Naurois}, {Niederwanger}, {Niemiec}, {Oakes},
  {O'Brien}, {Odaka}, {{\"O}ttl}, {Ohm}, {de O{\~n}a Wilhelmi}, {Ostrowski},
  {Oya}, {Padovani}, {Panter}, {Parsons}, {Paz Arribas}, {Pekeur}, {Pelletier},
  {Perennes}, {Petrucci}, {Peyaud}, {Pita}, {Poon}, {Prokhorov}, {Prokoph},
  {P{\"u}hlhofer}, {Punch}, {Quirrenbach}, {Raab}, {Reimer}, {Reimer},
  {Renaud}, {de los Reyes}, {Rieger}, {Romoli}, {Rosier-Lees}, {Rowell},
  {Rudak}, {Rulten}, {Sahakian}, {Salek}, {Sanchez}, {Santangelo}, {Sasaki},
  {Schlickeiser}, {Sch{\"u}ssler}, {Schulz}, {Schwanke}, {Schwemmer},
  {Settimo}, {Seyffert}, {Shafi}, {Shilon}, {Simoni}, {Sol}, {Spanier},
  {Spengler}, {Spies}, {Stawarz}, {Steenkamp}, {Stegmann}, {Stinzing}, {Stycz},
  {Sushch}, {Tavernet}, {Tavernier}, {Taylor}, {Terrier}, {Tibaldo}, {Tiziani},
  {Tluczykont}, {Trichard}, {Tuffs}, {Uchiyama}, {Valerius}, {van der Walt},
  {van Eldik}, {van Soelen}, {Vasileiadis}, {Veh}, {Venter}, {Viana},
  {Vincent}, {Vink}, {Voisin}, {V{\"o}lk}, {Vuillaume}, {Wadiasingh}, {Wagner},
  {Wagner}, {Wagner}, {White}, {Wierzcholska}, {Willmann}, {W{\"o}rnlein},
  {Wouters}, {Yang}, {Zabalza}, {Zaborov}, {Zacharias}, {Zdziarski}, {Zech},
  {Zefi}, {Ziegler}, \& {{\.Z}ywucka}}]{2018A&A...612A...2H}
{H.~E.~S.~S. Collaboration}, {Abdalla}, H., {Abramowski}, A., {et~al.} 2018,
  \aap, 612, A2, \dodoi{10.1051/0004-6361/201629377}

\bibitem[{{Izawa} {et~al.}(2015){Izawa}, {Dotani}, {Fujinaga}, {Bamba},
  {Ozaki}, \& {Hiraga}}]{2015PASJ...67...43I}
{Izawa}, M., {Dotani}, T., {Fujinaga}, T., {et~al.} 2015, \pasj, 67, 43,
  \dodoi{10.1093/pasj/psv013}

\bibitem[{{Kirichenko} {et~al.}(2015){Kirichenko}, {Shibanov}, {Shternin},
  {Johnston}, {Voronkov}, {Danilenko}, {Barsukov}, {Lai}, \&
  {Zyuzin}}]{2015MNRAS.452.3273K}
{Kirichenko}, A., {Shibanov}, Y., {Shternin}, P., {et~al.} 2015, \mnras, 452,
  3273, \dodoi{10.1093/mnras/stv1420}

\bibitem[{{Lemoine-Goumard} {et~al.}(2011){Lemoine-Goumard}, {Zavlin},
  {Grondin}, {Shannon}, {Smith}, {Burgay}, {Camilo}, {Cohen-Tanugi}, {Freire},
  {Grove}, {Guillemot}, {Johnston}, {Keith}, {Kramer}, {Manchester},
  {Michelson}, {Parent}, {Possenti}, {Ray}, {Renaud}, {Thorsett}, {Weltevrede},
  \& {Wolff}}]{2011A&A...533A.102L}
{Lemoine-Goumard}, M., {Zavlin}, V.~E., {Grondin}, M.~H., {et~al.} 2011, \aap,
  533, A102, \dodoi{10.1051/0004-6361/201117413}

\bibitem[{{Mattana} {et~al.}(2009){Mattana}, {Falanga}, {G{\"o}tz}, {Terrier},
  {Esposito}, {Pellizzoni}, {De Luca}, {Marandon}, {Goldwurm}, \&
  {Caraveo}}]{2009ApJ...694...12M}
{Mattana}, F., {Falanga}, M., {G{\"o}tz}, D., {et~al.} 2009, \apj, 694, 12,
  \dodoi{10.1088/0004-637X/694/1/12}

\bibitem[{{Murphy} {et~al.}(2007){Murphy}, {Mauch}, {Green}, {Hunstead},
  {Piestrzynska}, {Kels}, \& {Sztajer}}]{2007yCat.8082....0M}
{Murphy}, T., {Mauch}, T., {Green}, A., {et~al.} 2007, VizieR Online Data
  Catalog, VIII/82

\bibitem[{{Renaud} {et~al.}(2008){Renaud}, {Hoppe}, {Komin}, {Moulin},
  {Marandon}, \& {Clapson}}]{2008AIPC.1085..285R}
{Renaud}, M., {Hoppe}, S., {Komin}, N., {et~al.} 2008, in American Institute of
  Physics Conference Series, Vol. 1085, American Institute of Physics
  Conference Series, ed. F.~A. {Aharonian}, W.~{Hofmann}, \& F.~{Rieger},
  285--288, \dodoi{10.1063/1.3076661}

\bibitem[{{Sironi} {et~al.}(2015){Sironi}, {Keshet}, \&
  {Lemoine}}]{2015SSRv..191..519S}
{Sironi}, L., {Keshet}, U., \& {Lemoine}, M. 2015, \ssr, 191, 519,
  \dodoi{10.1007/s11214-015-0181-8}

\bibitem[{{Tibaldo} {et~al.}(2018){Tibaldo}, {Zanin}, {Faggioli}, {Ballet},
  {Grondin}, {Hinton}, \& {Lemoine-Goumard}}]{2018A&A...617A..78T}
{Tibaldo}, L., {Zanin}, R., {Faggioli}, G., {et~al.} 2018, \aap, 617, A78,
  \dodoi{10.1051/0004-6361/201833356}

\bibitem[{{Torres} {et~al.}(2014){Torres}, {Cillis}, {Mart{\'\i}n}, \& {de
  O{\~n}a Wilhelmi}}]{2014JHEAp...1...31T}
{Torres}, D.~F., {Cillis}, A., {Mart{\'\i}n}, J., \& {de O{\~n}a Wilhelmi}, E.
  2014, Journal of High Energy Astrophysics, 1, 31,
  \dodoi{10.1016/j.jheap.2014.02.001}

\bibitem[{{Wood} {et~al.}(2017){Wood}, {Caputo}, {Charles}, {Di Mauro},
  {Magill}, {Perkins}, \& {Fermi-LAT Collaboration}}]{2017ICRC...35..824W}
{Wood}, M., {Caputo}, R., {Charles}, E., {et~al.} 2017, in International Cosmic
  Ray Conference, Vol. 301, 35th International Cosmic Ray Conference
  (ICRC2017), 824.
\newblock \doarXiv{1707.09551}

\bibitem[{{Xin} {et~al.}(2018){Xin}, {Liao}, {Guo}, {Yuan}, {Liu}, {Fan}, \&
  {Wei}}]{2018ApJ...867...55X}
{Xin}, Y.-L., {Liao}, N.-H., {Guo}, X.-L., {et~al.} 2018, \apj, 867, 55,
  \dodoi{10.3847/1538-4357/aae313}

\bibitem[{{Zabalza}(2015)}]{2015ICRC...34..922Z}
{Zabalza}, V. 2015, in International Cosmic Ray Conference, Vol.~34, 34th
  International Cosmic Ray Conference (ICRC2015), 922.
\newblock \doarXiv{1509.03319}

\bibitem[{{Zavlin}(2007)}]{2007ApJ...665L.143Z}
{Zavlin}, V.~E. 2007, \apjl, 665, L143, \dodoi{10.1086/521300}

\end{thebibliography}
\bibliographystyle{aasjournal.bst}

\end{document}